\documentclass[twocolumn,usenatbib]{aa}

\usepackage{graphicx,rotating,portland,fancyheadings,natbib,amsmath}
\usepackage{txfonts}
\usepackage{longtable,lscape}
\def\aro{{$\alpha_{\rm ro}$~}}
\def\aox{{$\alpha_{\rm ox}$~}}

\def\ergas{{erg~s$^{-1}$~}}

\newcommand{\lsim}{{\lower.5ex\hbox{$\; \buildrel < \over \sim \;$}}}
\newcommand{\gsim}{{\lower.5ex\hbox{$\; \buildrel > \over \sim \;$}}}

\def\fxfr{$f_{\rm x}/f_{\rm r}$~}

\def\ee{\end{equation}}
\def\be{\begin{equation}}

\begin{document}

\title{ {\it ROXA}: a new multi-frequency selected large sample of blazars with SDSS and 2dF optical spectroscopy}
\author{Sara~Turriziani\inst{1}$^,$\inst{2},
 Elisabetta~Cavazzuti\inst{2}$^,$\inst{3}
and Paolo~Giommi\inst{2}$^,$\inst{3}
\institute{  Universit\`a degli studi di Roma, Tor Vergata, Dip. Fisica,
             via della ricerca scientifica 1,  00133  Roma, Italy.
\and
            ASI Science Data Center, ASDC c/o ESRIN,
            via G. Galilei 00044 Frascati, Italy.
\and
            Agenzia Spaziale Italiana,
            Unit\`a Osservazione dell'Universo,
            viale Liegi 26 00198 Roma, Italy
}}
\offprints{S. Turriziani, turriziani@asdc.asi.it}
\date{Received ....; Accepted ....}

\authorrunning{Turriziani et al.}
    \titlerunning{ {\it ROXA} MultiFrequecy Blazar Sample}

\abstract
{ Although Blazars are a small fraction of the overall AGN population they are expected to be the dominant population of extragalactic sources in the hard X-ray and gamma-ray bands and have been shown to be the largest contaminant of CMB fluctuation maps. So far the number of known blazars is of the order of several hundreds, but  the forthcoming AGILE, GLAST and Planck space observatories will detect several thousand of  objects of this type.}
{In preparation for these missions it is necessary to identify new samples of blazars to study their multi-frequency characteristics and statistical properties.}
{ We compiled a sample of objects with blazar-like properties via a cross-correlation between large radio (NVSS, ATCAPMN) and X-ray surveys (RASS) using the SDSS-DR4 and 2dF survey data to spectroscopically identify our candidates and test the validity of the selection method. }
{ We present the Radio - Optical - X-ray catalog built at ASDC  ({\it ROXA}), a list of 816 objects among which 510 are confirmed blazars. 
Only 19\% of the candidates turned out to be certainly non-blazars demonstrating the high efficiency of our selection method. }
{Our catalog includes 173 new blazar identifications, or about 10\% of all presently known blazars. The relatively high flux threshold in the X-ray energy band (given by the RASS survey) preferentially selects objects with high \fxfr ratio leading to the discovery of new High Energy Peaked BL Lac (HBLs). 
Our catalog therefore includes many new potential targets for GeV-TeV observations.
}


\maketitle

\section{Introduction}

Blazars are the rarest ($ \sim 5 \%$) and most extreme type of Active Galactive Nuclei (AGN) known. 
Historically, the classification of AGN has been largely based on observational characteristics leading to the proliferation of different classes of objects. All sources of this type, however, can be seen as part of a general  paradigm in which AGN are divided into Thermal Emission Dominated (TED) AGN, where the emitted radiation is mostly generated  through the accretion process onto a super-massive black hole, and Non-Thermal Emission Dominated (NTED) AGN, where  the observed emission is mostly non-thermal and is generated in a jet of material moving away from the nucleus at relativistic speeds  \cite[e.g.][]{giocola06}. Within this framework blazars are the small subset of NTED AGN in which the jet is closely aligned to the line of sight  causing their emission to be strongly amplified by relativistic effects \citep[as originally proposed by][]{Bla78}.  

We recognize a source as a blazar if it shows the properties usually associated to aligned beamed emission such as strong and rapidly variable emission in all energy bands, from radio to GeV, sometimes TeV energies, core dominated radio emission with flat radio spectral index, superluminal motion of radio compact regions, the presence of one sided jets (a jet on the other side is thought to exist, but with emission that is de-amplified by relativistic effects) and high brightness temperatures ($T_b \sim 10^{11} - 10^{18}K$), close to or above the Compton limit ($T_b \approx 10^{12}$). 

Sources that initially show only some of these properties, in later observations often also show the others, strengthening the hypothesis that these are basically equivalent and inseparable features related to the same underlying physical process. \\

Blazars  include  BL Lacertae objects (BL Lacs) where we observe a non-thermal optical continuum with no or very weak emission lines, and Flat Spectrum Radio Quasars (FSRQs) which exhibit both strong narrow and broad emission lines.
Furthermore, BL Lacs can be  distinguished by the peak of the synchrotron emission in their Spectral Energy Distributions (SEDs). Objects with synchrotron peak at low energy (typically in the Infra-Red) are generally found in radio surveys and are called {\bf LBLs} (Low Energy Peaked BL Lacs), whereas the rarer objects with synchrotron peak in the UV/X-ray band are called {\bf HBLs} (High Energy Peaked BL Lacs) and  are mainly selected in the X-ray band, where the maximum of their synchrotron power is emitted \citep[e.g.][]{P95}.\\

In this paper we discuss a large sample of  {\it candidate blazars} that we have assembled using a 
multi-frequency selection technique based on the NVSS \citep{Con98}, ATCAPMN (ATCA catalogue of compact PMN sources in \cite{Tasker}, RASS \citep{Vog99,Vog00} 
and GSC2 \citep{lasker95,McLean00} catalogs.

We have assessed the quality of the sample using a subsample of 816 objects for which data from Sloan Digital Sky Survey - Data Release 4 \citep[SDSS-DR4;][]{sdssdr4}, 2dF Galaxy Redshift Survey \citep[2dFGRS;][]{2dFGRS} and 2dF QSO Survey \citep[2dFQSO;][]{2dFQSO} are available.

We refer to our final list of 816 sources as the ROXA catalog, which stands for Radio Optical X-ray ASDC catalog.

We have built {\it ROXA} as a  complement to the work of \cite{SE03, SE04, SE05} as we share their goal of significantly enlarging the existing blazar catalogs in order to understand blazars statistical properties and to select interesting objects for upcoming high energy astronomy missions like AGILE, and GLAST. \cite{SE05},  with a single-band approach, selected flat-spectrum radio sources and then used optical telescopes for spectroscopic follow-up observations, whereas in this work we use a multi-frequency approach that requires no additional telescope time.  Moreover, the selection of radio sources results mostly in the discovery of LBL sources whereas our flux limits, which are mostly driven by the relatively shallow X-ray sensitivity of the RASS survey, favor objects such as HBL since they emit the maximum 
of their synchrotron emission near or within the X-ray band. 
This bias towards HBL sources is strengthened by the Ultra-Violet excess requirement in the SDSS spectroscopical target selection \citep{rich2002}. The presence of HBLs allows {\it ROXA} to contribute with new targets for GeV/TeV observations. \\
  
 {\it ROXA} is also an augmentation of \cite{Collinge05} who searched for extragalatic quasi-featurless objects in the Second Data Release of Sloan Digital Sky Survey \citep[SDSS-DR2;][]{sdssdr2}  with low proper motion in USNO-B Catalog \citep{USNOB} to built a BL Lac candidate sample. However, further information in other bands is required to confirm classifications therefore we used a multifrequency selection algorithm to build the candidate blazar sample and, although radio and X-ray selections could introduce biases, the high efficiency of the selection method allowed us to find many new blazars. \\
 
 The blazar selection algorithm is described in Section 2. In section 3 we focus on optical counterpart investigation and source classification criteria. Finally, we discuss our results and present our future projects in Section 4. \\

Throughout the paper we assume a cosmology with parameters in agreement with the results of Wilkinson Microwave Anisotropy Probe \citep[WMAP;][]{bennett03a}: $H_{0}$=70 km s$^{-1}$ Mpc$^{-1}$,~$\Omega_{M}$=0.3, $\Omega_{\Lambda}$=0.7 \citep{cosmopar03}. \\

\section{The selection method}

For the definition of our sample we relied on the availability of large catalogs of astronomical objects combined with on-line services offering simple access to finding charts and magnitude estimates. 

The method is similar to that used for the DXRBS \citep{L01,Pad06} and the Sedentary Survey \citep{Gio99,Gio05} and consists in three steps: 
\begin{enumerate} 
\item a first cross-correlation between radio and X-ray surveys (the NRAO VLA Sky Survey \citep[NVSS;][]{Con98}, ATCAPMN (ATCA catalogue of compact PMN sources in \cite{Tasker}) and ROSAT All Sky Survey \citep[RASS;][]{Vog99,Vog00}). 
\item
For each radio/X-ray match, optical magnitudes were retrieved from the Guide Star Catalog \citep[GSC2;][]{lasker95,McLean00} 

\item For all radio/optical/X-ray matches we calculated the X-ray to optical (\aox) and radio to optical (\aro) spectral slopes and took only sources with \aox and \aro values within the blazar area \citep[for more detail see e.g.][]{Per98,Gio99,L01,Pad06}. 
 \end{enumerate}

\subsection{The X-ray - radio cross correlation}

The NVSS catalog of radio sources includes nearly two million objects above a flux limit of 2.5 mJy at 1.4GHz \citep{Con98}, the ATCAPMN catalog includes 7178 objects and it is reasonibly complete down to $\sim$ 100 mJy \citep{Tasker} whereas the RASS X-ray catalog is a list of 124.735 objects detected during ROSAT All Sky Survey \citep{Vog99,Vog00}. \\
As for the case of the Sedentary Survey \citep{Gio99} and the DXRBS \citep{Pad06}, we used the 
EXOSAT/BROWSE software to cross-correlate the NVSS and the RASS catalogs. The cross-correlation used a radius of 1 arcmin and resulted in 16.596 matches. \\
We then restricted our sample to sources with Galactic latitude $\left\lvert\ b \right\rvert  > 20^{o}$  finding 12.988 candidates. We kept only those with $\Delta_{rx} < 2.5\sigma_{rx}$, where $\sigma_{rx} = \sqrt{ \sigma_{x}^2 + \sigma_{r}^2} $ is the total (radio+X-ray) positional uncertainty and $\Delta_{rx}$ is the actual distance between the radio and X-ray positions. This resulted in 9663 entries.  \\
 
We estimated the percentage of accidental spatial coincidences of unrelated objects by shifting the coordinates of all the sources in one of the catalogs by a fixed amount and re-running the cross-correlation with the same parameters. This procedure led to the statistical estimation that only 1 - 2 \% could be spurious associations. \\

\subsection{ Blazar candidate selection}

For each RASS-NVSS match we obtained the magnitudes of the optical candidate counterparts from the GSC2 catalog. When no GSC2 source was present within the few arc-seconds positional uncertainty of the NVSS radio source, we assumed that the counterpart is fainter than Jmag=19.5, approximately the flux limit of GSC2 \citep{McLean00}.
In the very rare cases when more than one GSC2 object was present, the optically brightest object was initially assumed to be the counterpart.  

We used EXOSAT/BROWSE software to get 5 GHz fluxes via a cross-correlation with several radio catalogs, such as NORTH6cm \citep{north6cm}, GB6 \citep{GB6} and  Parkes-MIT-NRAO \citep[PMN;][]{Gri93}. For sources with no 5 GHz flux information we estimated it as $F_{5 Ghz} = F_{1.4 GHz} \times (\frac{20}{6})^{-\alpha_{rm}}$, where $\alpha_{rm}$ is 0.25, the assumed average value for flat spectrum sources. \\

We corrected X-ray and optical fluxes for Galactic absorption and the redshift-dependent K-correction. For sources with no redshift information we assumed a typical redshift equal to the average value found in sources with the same X-ray to radio flux ratio (\fxfr). 

To build our candidate blazar sample we have calculated the optical to X-ray (\aox) and 
radio to optical (\aro) spectral slopes and selected objects with \aro  $>$ 0.2  where most of the blazars are located in the \aox-\aro diagram  \citep[see e.g.][for details]{Gio99}.
 
We relaxed the radio-loudness condition \aro $>$ 0.2 to \aro $>$ 0.1 (where \aro and \aox are the usual effective spectral indices defined between the rest-frame frequencies of 5 GHz, 5000 $\AA$ and 1 keV) for objects which had no optical counterpart in the GSC2, since a lower limit on Jmag results in a lower limit on \aro. \\

\begin{figure}[ht]
\vbox{
\centerline{
\includegraphics[width=10.0cm, angle=0] {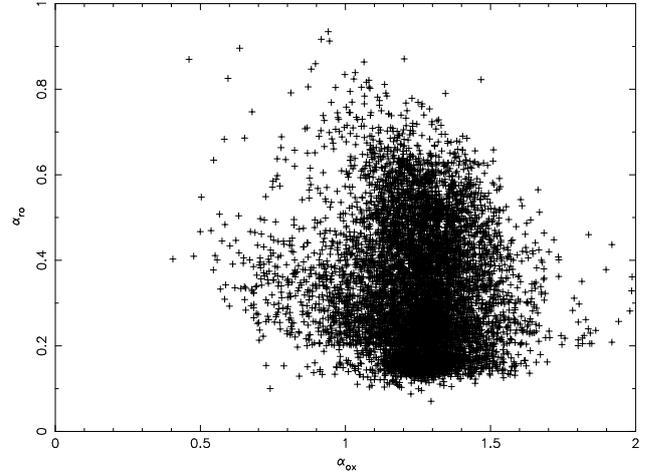}
}}
\caption{The \aox -- \aro distribution for the whole sample of 7.662 blazar candidates}
\label{aroaox_all}
\end{figure}

\begin{figure}[ht]
\vspace{-4.cm}
\vbox{
\centerline{
\includegraphics[width=10.8cm, angle=0] {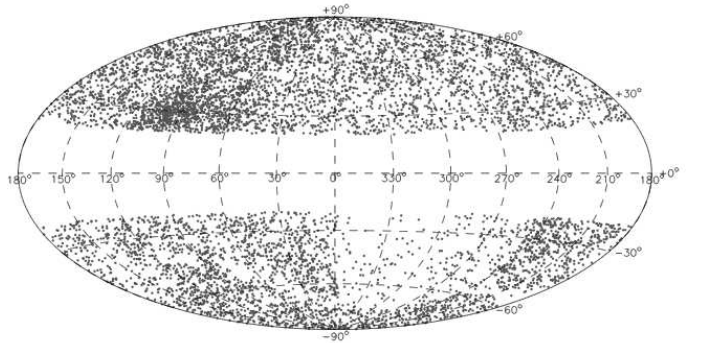}
}}
\vspace{-4.0cm}
\caption{The Galactic coordinates of the 7662 blazar candidates are plotted in Aitoff projection}
\label{aitoff_blcand}
\end{figure}

The selection process led to a sample of 7662 blazar candidates.  Fig. \ref{aroaox_all} shows the \aro- \aox diagram and Fig. \ref{aitoff_blcand} plots all 7662 sources in Galactic coordinates. The non-uniform space distribution in this last plot reflects the sensitivity of the RASS survey, which was higher around the ecliptic coordinates, and the higher radio flux limit (hence lower source density) of the ATCAPMN catalog at declinations south of -40$^o$. 

\section{Classification of optical counterparts using SDSS and 2dF data}

For all sources in our sample we used the SDSS on-line services (SkyServer-Radial search)  to download SDSS measurements (right ascension, declination and {\it u' g' r' i' z'} magnitudes) of each optical object  lying within 6 arcseconds from each NVSS radio position.

We found that 2.353 RASS+NVSS sources have SDSS-DR4 optical counterparts (including some multiple optical counterparts, MOCs ). \\

We then used the SkyServer DR4 Spectroscopic Query (proximity with radius 10 arcseconds) and the DAS online services to acquire spectra.  We used previous downloaded SDSS positions to get spectral parameters and survey files for objects classified as galaxies, stars or "unknown" in the SDSS standard products. We found 669 spectra among the 2353 SDSS optical counterparts. 

In a similar way we searched for optical counterparts also in the 2dFGRS and in the 2dFQSO databases. This resulted in spectra and $B_j$ magnitude estimates for additional 199 objects from 2dFGRS and for 94 from 2dFQSO (including MOCs). 

\subsection{Sources Classification}

All candidates that had MOCs were associated to a single optical object through a visual inspection procedure using the NVSS, ESO and NED online services. For each source we cross-checked the optical (ESO and SuperCOSMOS Surveys) finding chart with the RASS and NVSS positions and with the position in the FIRST catalog \citep{White97} when available. In this way we were able to chose the best optical counterpart for each blazar candidate. Moreover, we excluded those objects with RASS source extent greater than 30 arcsec. We did not apply this condition to those detections with 20 or less photons since in these cases the uncertainty in the estimation of X-ray extent is too large.

We built SEDs and analyzed optical spectra, with CaH\&K evaluation when needed, to classify each object (the equivalent width of emission lines in radio-loud AGN depends strongly on the contribution from non-thermal jet emission, and, therefore, on the Ca $H\&K$ break value \citep{Marcha96,Sca97}). Sources with SDSS magnitude estimates, although not targeted for spectroscopy by SDSS pipelines, were kept in the sample only if already classified in the literature as reported in NED. \\

SDSS and 2dF automated redshift estimation and spectral classification proved to be unreliable for several BL Lac objects because of the lack of emission lines. When possible, we re-calculated redshifts  using the IRAF analysis package \citep{iraf1,iraf2}. \\

We also defined a {\it transition class} for those objects that show properties in between two {\it standard} classifications. 
Regarding Radio Galaxies and BL Lacs we defined sources to be :

\begin{itemize}
\item BL Lac if  $L_x > 10^{44}$ \ergas or CaH\&K $<$ 0.4 or both
\item Radio Galaxy if CaH\&K $>$ 0.4 AND $L_x < 5 \times 10^{43}$ \ergas
\item Radio Galaxy/BL Lac transition object if CaH\&K $>$ 0.4 AND $L_x$ between $5 \times 10^{43}$ and
$10^{44}$ \ergas
\end{itemize}

Moreover we identified other {\it transition classes} for those objects showing optical properties which are borderline for their {\it standard} classification. These are BL Lac/FSRQ transition object and R.G./FSRQ transition object. 

We re-calculated \aro and \aox parameters estimating Vmag as in the following:
\begin{itemize}
\item for SDSS objects: V = {\it g'} - 0.55({\it g'} - {\it r'}) - 0.03
\item for 2dF objects: V = $B_j$ + 0.5
\item for object detected by both SDSS and 2dF we used SDSS magnitudes.
\end{itemize}

Our final catalog {\it ROXA} includes 816 objects,  510 of which are confirmed blazars (i.e.  62.5\%).  
Of the remaining sources 110 (or 13\% of the total) are confirmed QSOs (by the SDSS or 2dF optical spectra) but their  radio spectral slope is currently not available and therefore they remain blazar candidates.  Only 19\% of the candidates turned out to be definitely non-blazars.

\begin{table*}
\begin{center}
\caption{ROXA Catalog Statistic}
\label{stat_roxa}
\vspace{0.5cm}
\begin{tabular}{lcccc}
\hline
\multicolumn{1}{l}{ }& \multicolumn{1}{c}{Known objects} &\multicolumn{1}{c} {new objects}  &
\multicolumn{1}{c} {Total} &\multicolumn{1}{c} {Sample \%}\\
\hline
BL LAC                           & 182 & 60  & 242 &  29.7\%\\
\hline
FSRQ                             & 147 & 99  & 246 & 30.1\%\\
\hline
BL Lac/FSRQ transition obj.      &  1 &  7 &   8 &  1.0\%\\
\hline
BL LAC candidate                 &  7  &  7  &  14 &  1.7\%\\
\hline
{\bf confirmed blazars}          &{\bf 337 } &{\bf 173 } &{\bf 510} &{\bf 62.5\%}\\
\hline
R.G./BL Lac transition obj.      &  0 &  24  &  24 &  3.0\%\\
\hline
R.G./FSRQ transition obj.        &  0 &  2  &   2 &  0.2\%\\
\hline
Radio Galaxies                   & 24 &  8 &  32 &  3.9\%\\
\hline
SSRQ                             & 59 & 43 & 102 & 12.5\%\\
\hline
QSO RL                           & 60 & 50 & 110 & 13.5\%\\
\hline
Galaxies NELG                    & 12 &  2 &  14 &  1.7\%\\
\hline
BLRG                             &  2  &  0 &  2 &  0.2\%\\
\hline
others                           & 18 &  2 &  20 &  2.5\%\\
\hline
Total                            & 512 & 304 & 816 & 100\%\\
\hline
\end{tabular}
\end{center}
\end{table*}


Table \ref{stat_roxa} summarizes the statistics of our source classification. The complete catalog is reported in Table \ref{roxa}. 
The first column of Table \ref{stat_roxa} contains source classification, the second column gives the number of previously known objects (that is sources previously reported in the literature as blazars), the third column gives the number of new objects (classified in this work for the first time), the fourth column gives the total number of objects and the last column gives the percentage with respect to the entire sample . 

The label {\bf confirmed blazars} includes BL LAC, FSRQ, BL Lac/FSRQ transition object and BL Lac candidate. 

The label ``others`` is used to comprise several non-blazar sources, such as stars and variuos types of galaxies (Normal, Starburst and Seyfert). \\
We point out that we identified 173 new blazars. \\

\begin{figure}[ht]
\hspace{0.5cm}
\vbox{
\centerline{
\includegraphics[width=11.0cm, angle=0] {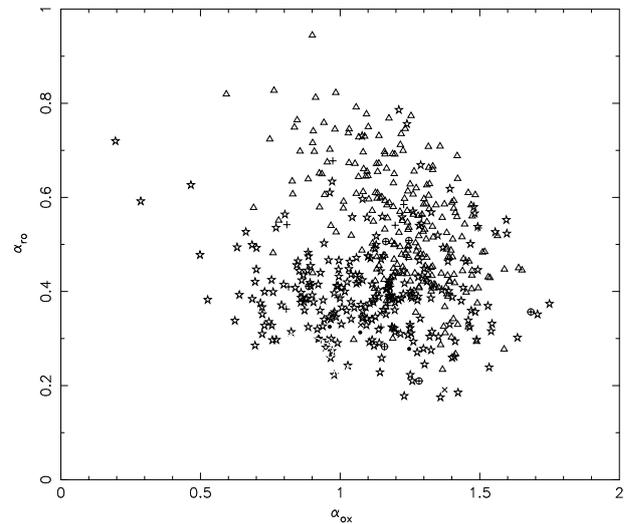}
}}

\caption{ROXA \aro -- \aox distribution of the confirmed blazars. Symbols are as follows: ${\star}$ -- BL Lac, ${\triangle}$ -- FSRQ, ${\bullet}$ -- HFSRQ, ${\oplus}$ -- BL Lac/FSRQ, + -- BL Lac candidate, x -- HFSRQ candidate}
\label{aro-aox-blazar}
\end{figure}

Fig. \ref{aro-aox-blazar} reports the multi-band indexes distribution of {\it ROXA} confirmed blazars.

\section{Conclusions and future work}
 
In this paper we presented the {\it ROXA} (Radio Optical X-ray ASDC)  catalog, a list of  of 816 objects including 510 confirmed blazars. 
This catalog has been extracted from a sample of 7662 objects selected, through the cross matching procedure between the NVSS, ATCAPMN, RASS and the GSC2 catalog, to have properties similar to those of previously known blazars in the \aro-\aox diagram.

We tested the selection method, also used in the DXRBS and in the Sedentary surveys, via a cross-correlation of the candidate sample with the SDSS DR4 and 2dF surveys for which optical spectra are available. We analyzed SEDs and spectra to classify the candidate blazars and we confirmed the validity of the selection algorithm.

Our final catalog consists of $\sim$ 62.5\% confirmed blazars, $\sim$ 13\% QSO with no radio spectral information and of $\sim$ 19\% of sources that are surely non blazars. Since the ratio between flat and steep spectrum QSO is 254/102 (see tab.  \ref{stat_roxa} ) the majority of QSO with no radio spectral information is expected to be blazars. We can therefore conclude that our selection algorithm has an efficiency of  ($ \sim 70\%$).
We also contributed several new blazar identifications finding 173 new blazars. 

Our method for the construction of the {\it ROXA} catalog is complementary to the work of \cite{SE05} and \cite{Collinge05} as these authors used single-band approaches whereas we used a multi-frequency selection algorithm. \cite{SE05} selected flat spectrum radio sources and then requested telescope time for follow-up spectroscopic observations whereas \cite{Collinge05} searched for extragalatic quasi-featurless objects in SDSS-DR2 with low proper motion in USNO-B Catalog to built a BL Lac candidate sample.  However further information in other bands are required to confirm the classifications. 
Moreover, the selection of radio sources results in finding mostly LBLs by \cite{SE05} whereas our algorithm have greater chances of finding HBLs, because of our relatively high X-Ray flux limit that preferentially select objects in which the maximum of synchrotron power is emitted at high energy. 

We share the goal with \cite{SE05} to create as large as possible blazar catalogs to be used as targets for observations during the next planned new high energy missions (e.g. AGILE and GLAST). \\

We plan to request high frequency radio observations in order to measure the spectral slope of the 110 QSOs with radio measurements at a single frequency and to better evaluate the nuclear spectra  of the 102 Steep Spectrum Radio Quasars (SSRQs) as their steep slopes could be due to extended radio components.

In principle, by applying this selection method to the entire sky and not just to  limited regions, we could build a much larger sample suitable for the identification of foreground sources in CMB maps and in gamma-ray data. At present, the most suitable optical data available are from the SDSS project. For this reason we intend to update the ROXA catalog as new SDSS releases become available.


\begin{acknowledgements}

 {\it  We are grateful to Silvia Piranomonte for her help in the analysis of part of the optical spectra. This work is partly based on data from the 2dF, SDSS, NVSS, ROSAT, GSC2, GB6,  Parkes-MIT-NRAO (PMN), NORTH6cm catalogs and services.
Additional data have been obtained from the NASA/IPAC Extragalactic Database (NED)  and from the ASI Science Data Center (ASDC). 
 
Funding for the creation and distribution of the SDSS Archive has been provided by the Alfred P. Sloan Foundation, the Participating Institutions, the National Aeronautics and Space Administration, the National Science Foundation, the U.S. Department of Energy, the Japanese Monbukagakusho, and the Max Planck Society. The SDSS is managed by the Astrophysical Research Consortium (ARC) for the Participating Institutions.

IRAF is distributed by the National Optical Astronomy Observatories,
    which are operated by the Association of Universities for Research
    in Astronomy, Inc., under cooperative agreement with the National
    Science Foundation
}

\end{acknowledgements}

\bibliography{ROXA_astroph}

\begin{thebibliography}{34}
\expandafter\ifx\csname natexlab\endcsname\relax\def\natexlab#1{#1}\fi

\bibitem[{{Abazajian} {et~al.}(2004){Abazajian}, {Adelman-McCarthy},
  {Ag{\"u}eros}, {Allam}, {Anderson}, {Anderson}, {Annis}, {Bahcall}, {Baldry},
  {Bastian}, {Berlind}, {Bernardi}, {Blanton}, {Bochanski}, {Boroski},
  {Briggs}, {Brinkmann}, {Brunner}, {Budav{\'a}ri}, {Carey}, {Carliles},
  {Castander}, {Connolly}, {Csabai}, {Doi}, {Dong}, {Eisenstein}, {Evans},
  {Fan}, {Finkbeiner}, {Friedman}, {Frieman}, {Fukugita}, {Gal}, {Gillespie},
  {Glazebrook}, {Gray}, {Grebel}, {Gunn}, {Gurbani}, {Hall}, {Hamabe},
  {Harris}, {Harris}, {Harvanek}, {Heckman}, {Hendry}, {Hennessy}, {Hindsley},
  {Hogan}, {Hogg}, {Holmgren}, {Ichikawa}, {Ichikawa}, {Ivezi{\'c}}, {Jester},
  {Johnston}, {Jorgensen}, {Kent}, {Kleinman}, {Knapp}, {Kniazev}, {Kron},
  {Krzesinski}, {Kunszt}, {Kuropatkin}, {Lamb}, {Lampeitl}, {Lee}, {Leger},
  {Li}, {Lin}, {Loh}, {Long}, {Loveday}, {Lupton}, {Malik}, {Margon},
  {Matsubara}, {McGehee}, {McKay}, {Meiksin}, {Munn}, {Nakajima}, {Nash},
  {Neilsen}, {Newberg}, {Newman}, {Nichol}, {Nicinski}, {Nieto-Santisteban},
  {Nitta}, {Okamura}, {O'Mullane}, {Ostriker}, {Owen}, {Padmanabhan},
  {Peoples}, {Pier}, {Pope}, {Quinn}, {Richards}, {Richmond}, {Rix}, {Rockosi},
  {Schlegel}, {Schneider}, {Scranton}, {Sekiguchi}, {Seljak}, {Sergey},
  {Sesar}, {Sheldon}, {Shimasaku}, {Siegmund}, {Silvestri}, {Smith}, {Smol{\v
  c}i{\'c}}, {Snedden}, {Stebbins}, {Stoughton}, {Strauss}, {SubbaRao},
  {Szalay}, {Szapudi}, {Szkody}, {Szokoly}, {Tegmark}, {Teodoro}, {Thakar},
  {Tremonti}, {Tucker}, {Uomoto}, {Vanden Berk}, {Vandenberg}, {Vogeley},
  {Voges}, {Vogt}, {Walkowicz}, {Wang}, {Weinberg}, {West}, {White}, {Wilhite},
  {Xu}, {Yanny}, {Yasuda}, {Yip}, {Yocum}, {York}, {Zehavi}, {Zibetti}, \&
  {Zucker}}]{sdssdr2}
{Abazajian}, K., {Adelman-McCarthy}, J.~K., {Ag{\"u}eros}, M.~A., {et~al.}
  2004, \aj, 128, 502

\bibitem[{{Adelman-McCarthy} {et~al.}(2006){Adelman-McCarthy}, {Ag{\"u}eros},
  {Allam}, {Anderson}, {Anderson}, {Annis}, {Bahcall}, {Baldry}, {Barentine},
  {Berlind}, {Bernardi}, {Blanton}, {Boroski}, {Brewington}, {Brinchmann},
  {Brinkmann}, {Brunner}, {Budav{\'a}ri}, {Carey}, {Carr}, {Castander},
  {Connolly}, {Csabai}, {Czarapata}, {Dalcanton}, {Doi}, {Dong}, {Eisenstein},
  {Evans}, {Fan}, {Finkbeiner}, {Friedman}, {Frieman}, {Fukugita}, {Gillespie},
  {Glazebrook}, {Gray}, {Grebel}, {Gunn}, {Gurbani}, {de Haas}, {Hall},
  {Harris}, {Harvanek}, {Hawley}, {Hayes}, {Hendry}, {Hennessy}, {Hindsley},
  {Hirata}, {Hogan}, {Hogg}, {Holmgren}, {Holtzman}, {Ichikawa}, {Ivezi{\'c}},
  {Jester}, {Johnston}, {Jorgensen}, {Juri{\'c}}, {Kent}, {Kleinman}, {Knapp},
  {Kniazev}, {Kron}, {Krzesinski}, {Kuropatkin}, {Lamb}, {Lampeitl}, {Lee},
  {Leger}, {Lin}, {Long}, {Loveday}, {Lupton}, {Margon},
  {Mart{\'{\i}}nez-Delgado}, {Mandelbaum}, {Matsubara}, {McGehee}, {McKay},
  {Meiksin}, {Munn}, {Nakajima}, {Nash}, {Neilsen}, {Newberg}, {Newman},
  {Nichol}, {Nicinski}, {Nieto-Santisteban}, {Nitta}, {O'Mullane}, {Okamura},
  {Owen}, {Padmanabhan}, {Pauls}, {Peoples}, {Pier}, {Pope}, {Pourbaix},
  {Quinn}, {Richards}, {Richmond}, {Rockosi}, {Schlegel}, {Schneider},
  {Schroeder}, {Scranton}, {Seljak}, {Sheldon}, {Shimasaku}, {Smith}, {Smol{\v
  c}i{\'c}}, {Snedden}, {Stoughton}, {Strauss}, {SubbaRao}, {Szalay},
  {Szapudi}, {Szkody}, {Tegmark}, {Thakar}, {Tucker}, {Uomoto}, {Vanden Berk},
  {Vandenberg}, {Vogeley}, {Voges}, {Vogt}, {Walkowicz}, {Weinberg}, {West},
  {White}, {Xu}, {Yanny}, {Yocum}, {York}, {Zehavi}, {Zibetti}, \&
  {Zucker}}]{sdssdr4}
{Adelman-McCarthy}, J.~K., {Ag{\"u}eros}, M.~A., {Allam}, S.~S., {et~al.} 2006,
  \apjs, 162, 38

\bibitem[{{Becker} {et~al.}(1991){Becker}, {White}, \& {Edwards}}]{north6cm}
{Becker}, R.~H., {White}, R.~L., \& {Edwards}, A.~L. 1991, \apjs, 75, 1

\bibitem[{Bennett {et~al.}(2003)Bennett, Bay, Halpern, Jackson, Jarosik, Kogut,
  Limon, Meyer, Page, Spergel, Tucker, Wilkinson, Wollack, \&
  Wright}]{bennett03a}
Bennett, C.~L., Bay, M., Halpern, M., {et~al.} 2003, ApJ, 583, 1

\bibitem[{Blandford \& Rees(1978)}]{Bla78}
Blandford, R.~D. \& Rees, M.~J. 1978, in Pittsburgh Conference on {BL} {L}ac
  Objects, ed. A.~M. Wolfe (University of Pittsburgh, Pittsburgh), 328

\bibitem[{{Colless} {et~al.}(2001){Colless}, {Dalton}, {Maddox}, {Sutherland},
  {Norberg}, {Cole}, {Bland-Hawthorn}, {Bridges}, {Cannon}, {Collins}, {Couch},
  {Cross}, {Deeley}, {De Propris}, {Driver}, {Efstathiou}, {Ellis}, {Frenk},
  {Glazebrook}, {Jackson}, {Lahav}, {Lewis}, {Lumsden}, {Madgwick}, {Peacock},
  {Peterson}, {Price}, {Seaborne}, \& {Taylor}}]{2dFGRS}
{Colless}, M., {Dalton}, G., {Maddox}, S., {et~al.} 2001, \mnras, 328, 1039

\bibitem[{Collinge {et~al.}(2005)Collinge, Strauss, Hall, Ivezic, Munn,
  Schlegel, Zakamska, Anderson, Harris, Richards, Schneider, Voges, York,
  Margon, \& Brinkmann}]{Collinge05}
Collinge, M., Strauss, M., Hall, P., {et~al.} 2005, AJ, aJ, in press

\bibitem[{Condon {et~al.}(1998)Condon, Cotton, Greisen, Yin, Perley, Taylor, \&
  Broderick}]{Con98}
Condon, J.~J., Cotton, W.~D., Greisen, E.~W., {et~al.} 1998, AJ, 115, 1693

\bibitem[{{Giommi} \& {Colafrancesco}(2006)}]{giocola06}
{Giommi}, P. \& {Colafrancesco}, S. 2006, ArXiv Astrophysics e-prints

\bibitem[{Giommi {et~al.}(1999)Giommi, Menna, \& Padovani}]{Gio99}
Giommi, P., Menna, M.~T., \& Padovani, P. 1999, MNRAS, 310, 465

\bibitem[{Giommi {et~al.}(2005)Giommi, Piranomonte, Perri, \& Padovani}]{Gio05}
Giommi, P., Piranomonte, S., Perri, M., \& Padovani, P. 2005, A\&A, 434, 385

\bibitem[{{Gregory} {et~al.}(1996){Gregory}, {Scott}, {Douglas}, \&
  {Condon}}]{GB6}
{Gregory}, P.~C., {Scott}, W.~K., {Douglas}, K., \& {Condon}, J.~J. 1996,
  \apjs, 103, 427

\bibitem[{Griffith \& Wright(1993)}]{Gri93}
Griffith, M.~R. \& Wright, A.~E. 1993, AJ, 105, 1666

\bibitem[{Landt {et~al.}(2001)Landt, Padovani, Perlman, Giommi, Bignall, \&
  Tzioumis}]{L01}
Landt, H., Padovani, P., Perlman, E.~S., {et~al.} 2001, MNRAS, 323, 757

\bibitem[{Lasker(1995)}]{lasker95}
Lasker, B.~M., e.~a. 1995, in ~in ESA SP-379, Future Possibilities for
  Astrometry in Space,, ed. v.~L. F. . G. T.-D. Perryman~M.A.C., 137

\bibitem[{March\~a {et~al.}(1996)March\~a, Browne, Impey, \& Smith}]{Marcha96}
March\~a, M. J.~M., Browne, I. W.~A., Impey, C.~D., \& Smith, P.~S. 1996,
  MNRAS, 281, 425

\bibitem[{McLean {et~al.}(2000)McLean, Greene, Lattanzi, \& Pirenne}]{McLean00}
McLean, B., Greene, G.~R., Lattanzi, M.~G., \& Pirenne, B. 2000, in ~in ASP
  Conf. Ser., ADASS IX, ed. V.~C. . C.~D. Manset~N., Vol. 216, 145--148

\bibitem[{{Monet} {et~al.}(2003){Monet}, {Levine}, {Canzian}, {Ables}, {Bird},
  {Dahn}, {Guetter}, {Harris}, {Henden}, {Leggett}, {Levison}, {Luginbuhl},
  {Martini}, {Monet}, {Munn}, {Pier}, {Rhodes}, {Riepe}, {Sell}, {Stone},
  {Vrba}, {Walker}, {Westerhout}, {Brucato}, {Reid}, {Schoening}, {Hartley},
  {Read}, \& {Tritton}}]{USNOB}
{Monet}, D.~G., {Levine}, S.~E., {Canzian}, B., {et~al.} 2003, \aj, 125, 984

\bibitem[{Padovani \& Giommi(1995)}]{P95}
Padovani, P. \& Giommi, P. 1995, ApJ, 444, 567

\bibitem[{Padovani {et~al.}(2006)Padovani, Giommi, Landt, \& Perlman}]{Pad06}
Padovani, P., Giommi, P., Landt, H., \& Perlman, E. 2006, in preparation

\bibitem[{Perlman {et~al.}(1998)Perlman, Padovani, Giommi, Sambruna, Jones,
  Tzioumis, \& Reynolds}]{Per98}
Perlman, E.~S., Padovani, P., Giommi, P., {et~al.} 1998, AJ, 115, 1253

\bibitem[{{Richards} {et~al.}(2002){Richards}, {Fan}, {Newberg}, {Strauss},
  {Vanden Berk}, {Schneider}, {Yanny}, {Boucher}, {Burles}, {Frieman}, {Gunn},
  {Hall}, {Ivezi{\'c}}, {Kent}, {Loveday}, {Lupton}, {Rockosi}, {Schlegel},
  {Stoughton}, {SubbaRao}, \& {York}}]{rich2002}
{Richards}, G.~T., {Fan}, X., {Newberg}, H.~J., {et~al.} 2002, \aj, 123, 2945

\bibitem[{R.L. {et~al.}(1997)R.L., R.H., D.J., \& M.D.}]{White97}
R.L., W., R.H., B., D.J., H., \& M.D., G. 1997, ApJ, 475, 479

\bibitem[{Scarpa \& Falomo(1997)}]{Sca97}
Scarpa, R. \& Falomo, R. 1997, A\&A, 325, 109

\bibitem[{{Shanks} {et~al.}(2000){Shanks}, {Boyle}, {Croom}, {Loaring},
  {Miller}, \& {Smith}}]{2dFQSO}
{Shanks}, T., {Boyle}, B.~J., {Croom}, S., {et~al.} 2000, in ASP Conf. Ser.
  200: Clustering at High Redshift, ed. A.~{Mazure}, O.~{Le F{\`e}vre}, \&
  V.~{Le Brun}, 57--+

\bibitem[{{Sowards-Emmerd} {et~al.}(2003){Sowards-Emmerd}, {Romani}, \&
  {Michelson}}]{SE03}
{Sowards-Emmerd}, D., {Romani}, R.~W., \& {Michelson}, P.~F. 2003, \apj, 590,
  109

\bibitem[{{Sowards-Emmerd} {et~al.}(2005){Sowards-Emmerd}, {Romani},
  {Michelson}, {Healey}, \& {Nolan}}]{SE05}
{Sowards-Emmerd}, D., {Romani}, R.~W., {Michelson}, P.~F., {Healey}, S.~E., \&
  {Nolan}, P.~L. 2005, \apj, 626, 95

\bibitem[{{Sowards-Emmerd} {et~al.}(2004){Sowards-Emmerd}, {Romani},
  {Michelson}, \& {Ulvestad}}]{SE04}
{Sowards-Emmerd}, D., {Romani}, R.~W., {Michelson}, P.~F., \& {Ulvestad}, J.~S.
  2004, \apj, 609, 564

\bibitem[{{Spergel} {et~al.}(2003){Spergel}, {Verde}, {Peiris}, {Komatsu},
  {Nolta}, {Bennett}, {Halpern}, {Hinshaw}, {Jarosik}, {Kogut}, {Limon},
  {Meyer}, {Page}, {Tucker}, {Weiland}, {Wollack}, \& {Wright}}]{cosmopar03}
{Spergel}, D.~N., {Verde}, L., {Peiris}, H.~V., {et~al.} 2003, \apjs, 148, 175

\bibitem[{Tasker(2000)}]{Tasker}
Tasker, N. 2000, Identifications from the PMN Southern Survey (Ph. D. Thesis)

\bibitem[{{Tody}(1986)}]{iraf1}
{Tody}, D. 1986, in Instrumentation in astronomy VI; Proceedings of the
  Meeting, Tucson, AZ, Mar. 4-8, 1986. Part 2 (A87-36376 15-35). Bellingham,
  WA, Society of Photo-Optical Instrumentation Engineers, 1986, p. 733., ed.
  D.~L. {Crawford}, 733--+

\bibitem[{{Tody}(1993)}]{iraf2}
{Tody}, D. 1993, in ASP Conf. Ser. 52: Astronomical Data Analysis Software and
  Systems II, ed. R.~J. {Hanisch}, R.~J.~V. {Brissenden}, \& J.~{Barnes},
  173--+

\bibitem[{{Voges} {et~al.}(2000){Voges}, {Aschenbach}, {Boller}, {Brauninger},
  {Briel}, {Burkert}, {Dennerl}, {Englhauser}, {Gruber}, {Haberl}, {Hartner},
  {Hasinger}, {Pfeffermann}, {Pietsch}, {Predehl}, {Schmitt}, {Trumper}, \&
  {Zimmermann}}]{Vog00}
{Voges}, W., {Aschenbach}, B., {Boller}, T., {et~al.} 2000, VizieR Online Data
  Catalog, 9029, 0

\bibitem[{Voges {et~al.}(1999)Voges, Aschenbach, Boller, Br\"auninger, Briel,
  Burkert, Dennerl, Englhauser, {et~al.}}]{Vog99}
Voges, W., Aschenbach, B., Boller, T., {et~al.} 1999, A\&A, 349, 389

\end{thebibliography}


\onecolumn
\scriptsize
\begin{landscape}
\begin{longtable}{lccccccccccl}
\caption{\label{roxa} The ROXA - Radio Optical X-ray ASDC Blazar Sample}\\
\hline
\multicolumn{1}{c}{Source name} &
\multicolumn{1}{c}{z} &
\multicolumn{1}{c}{$B_j$ mag} &
\multicolumn{1}{c}{{\it g'} mag} &
\multicolumn{1}{c}{Radio flux} &
\multicolumn{1}{c}{Radio flux} &
\multicolumn{1}{c}{X-ray flux} &
\multicolumn{1}{c}{$f_\mathrm{x}/f_\mathrm{r}$} &
\multicolumn{1}{c}{$L_x$} &
\multicolumn{1}{c}{$L_r$} &
\multicolumn{1}{c}{CaH\&K} &
\multicolumn{1}{c}{Classification$^{\mathrm{a}}$} \\
&  & & &1.4 GHz & 5 GHz&                      &                              &           &  & &\\
&  & & &mJy    & mJy  & erg $cm^{-2} s^{-1}$ & erg $cm^{-2} s^{-1} Jy^{-1}$ &erg $s^{-1}$& erg $s^{-1} Hz^{-1}$ & &\\
\hline
\endfirsthead

\caption{continued}\\
\hline
\multicolumn{1}{c}{Source name} &
\multicolumn{1}{c}{z} &
\multicolumn{1}{c}{$B_j$ mag} &
\multicolumn{1}{c}{{\it g'} mag} &
\multicolumn{1}{c}{Radio flux} &
\multicolumn{1}{c}{Radio flux} &
\multicolumn{1}{c}{X-ray flux} &
\multicolumn{1}{c}{$f_\mathrm{x}/f_\mathrm{r}$} &
\multicolumn{1}{c}{$L_x$} &
\multicolumn{1}{c}{$L_r$} &
\multicolumn{1}{c}{CaH\&K} &
\multicolumn{1}{c}{Classification$^{\mathrm{a}}$} \\
&  & & &1.4 GHz & 5 GHz&                      &                              &           &  & &\\
&  & & &mJy    & mJy  & erg $cm^{-2} s^{-1}$ & erg $cm^{-2} s^{-1} Jy^{-1}$ &erg $s^{-1}$& erg $s^{-1} Hz^{-1}$ & &\\
\hline
\endhead

\hline
\endfoot

ROXA J000132.8+145608.0&0.40&  0.0& 19.1&  315.4&  156.0& 9.90E-13&  6.35E-12&  5.55E+44&  1.77E+33& &SSRQ*\\                        
ROXA J000345.1-110818.4&1.57&  0.0& 20.0&  354.1&  118.0& 3.70E-13&  3.14E-12&  5.90E+45&  5.64E+34& &SSRQ*\\                        
ROXA J000416.6-290235.0&0.55& 18.2&  0.0&   16.1&    0.0& 1.20E-12&  1.01E-10&  1.47E+45&  1.97E+32& &QSO RL\\                        
ROXA J000558.4-275858.1&0.62& 18.0&  0.0&  309.1&    0.0& 1.00E-12&  4.37E-12&  1.65E+45&  5.09E+33& &FSRQ\\                        
ROXA J000559.2+160948.9&0.45&  0.0& 15.4&  805.3&    0.0& 8.70E-12&  1.46E-11&  6.49E+45&  6.01E+33& &SSRQ\\                        
ROXA J000622.6-000424.3&1.04&  0.0& 20.0& 3898.2&    0.0& 6.20E-13&  2.15E-13&  3.54E+45&  2.23E+35& &SSRQ\\                        
ROXA J001130.4+005751.8&1.49&  0.0& 20.4&  167.3&  140.0& 5.40E-13&  3.86E-12&  7.59E+45&  2.35E+34& &FSRQ\\                        
ROXA J001339.1-322443.8&0.26& 18.5&  0.0&  155.7&    0.0& 4.70E-13&  4.08E-12&  9.76E+43&  3.23E+32& &R.G./BL Lac*\\
\end{longtable}
\begin{list}{}{}
\item[$^{\mathrm{a}}$] *: new classification for those objects differently classified in literature; @: new object (never identified before this work); {$^z$}: objects with newly estimated redshifts 
\end{list}
\end{landscape}

\end{document}